\documentclass[copyright,creativecommons]{eptcs}
\providecommand{\event}{QPL 2019} 
\usepackage{breakurl}             
\usepackage{underscore}           

\usepackage{booktabs} 
\usepackage[utf8]{inputenc}
\usepackage[T1]{fontenc}
\usepackage{dsfont}

\usepackage[caption=no]{subfig}
\usepackage[linesnumbered]{algorithm2e}
\usepackage{tikz}
\usetikzlibrary{calc,chains,decorations.markings,matrix,patterns,positioning,shapes.geometric}
\usepackage{pgfplots, pgfplotstable}
\usepackage{mathtools}
\usepackage[noend]{algpseudocode}
\usepackage{array}
\usepackage{tabularx,amsmath,amssymb}
\usepackage{graphicx}
\usepackage{amsmath}
\usepackage{booktabs}
\newcolumntype{Z}{>{\raggedleft\arraybackslash}X}

\tikzset{flowbox/.style={draw,inner sep=2pt,align=center},
         flowtool/.style={inner sep=2pt,font=\footnotesize\itshape,align=center}}

\algdef{SE}[DOWHILE]{Do}{doWhile}{\algorithmicdo}[1]{\algorithmicwhile\ #1}%
\makeatletter
\newcommand{\nosemic}{\renewcommand{\@endalgocfline}{\relax}}
\newcommand{\dosemic}{\renewcommand{\@endalgocfline}{\algocf@endline}}
\makeatother

\SetKwInOut{Input}{Input}
\SetKwInOut{Output}{Output}

\newtheorem{example}{Example}

\newcommand{\cutimage}{} 
\newcommand{\B}{\ensuremath{\mathds{B}}}

\title{ROS: Resource-constrained Oracle Synthesis for Quantum Computers}
\author{Giulia Meuli \qquad \qquad Mathias Soeken
\institute{EPFL\\ Lausanne, Switzerland}
\email{ giulia.meuli@epfl.ch\qquad\qquad\qquad\qquad\qquad}
\and
Martin Roetteler
\institute{Microsoft Quantum\\ Redmond, United States}
\and
Giovanni De Micheli
\institute{EPFL\\ Lausanne, Switzerland}
}
\def\titlerunning{ROS: Resource-constrained Oracle Synthesis for Quantum Computers}
\def\authorrunning{G. Meuli, M. Soeken, M. Roetteler, G. De Micheli}

\begin{document}

%

\maketitle

\begin{abstract}
We present a completely automatic synthesis framework for oracle functions---a central part in many quantum algorithms. 
The proposed framework for resource-constrained oracle synthesis (\emph{ROS}) is a LUT-based hierarchical method in which every step is specifically tailored to address hardware resource constraints. 
\emph{ROS} embeds a LUT mapper designed to simplify the successive synthesis steps, costing each LUT according to the resources used by its corresponding quantum circuit. 
In addition, the framework exploits a SAT-based quantum garbage management technique. 
Those two characteristics give \emph{ROS} the ability to beat the state-of-the-art hierarchical method both in number of qubits and in number of operations.
The efficiency of the framework is demonstrated by synthesizing quantum oracles for Grover's algorithm.
\end{abstract}

%
%
%


\section{Introduction}
Practical quantum computers are nowadays a realistic pros\-pect thanks to advances in fabrication technology and the effort of the research community to revolutionize computing~\cite{debnath2016, MBK+16, MMS+16}. 
Quantum systems enable computation over superposition of states and are based on physical phenomena that are fundamentally different from the ones exploited in classical computing systems. For this reason, they require the development of dedicated logic synthesis tools.

The peculiarities of quantum computation can be exploited to solve problems that cannot be solved with standard computers in a reasonable time, by running innovative quantum algorithms. The possible applications span among others factorization~\cite{Shor94}, quantum chemistry~\cite{babbush16}, and satisfiability solving~\cite{Grover96}. 

Many quantum algorithms include combinational logic operations.
The large amount of resources necessary to perform such computations can overcome the resources available, hence preventing some algorithms to be computed on a constrained quantum hardware. Consequently, there is a large interest in finding synthesis methods that minimize the impact of combinational logic on the cost of quantum algorithms. 

Some automatic quantum circuit synthesis methods have been proposed~\cite{miller03, grosse09, amy17, schuch03}, which can be applied to relatively small logic designs. Hierarchical methods proved to be applicable to larger designs, as they are based on multi-level logic representations~\cite{rawski15}.
Among them, the LUT-based hierarchical reversible logic synthesis (\emph{LHRS}) framework has been proposed in~\cite{Soeken18}, and is currently part of the open source project \emph{RevKit}.\footnote{https://github.com/msoeken/revkit} It exploits classical logic synthesis methods to create quantum circuits of any given combinational logic component. Objective functions of the synthesis are: the number of qubits and the number of operations required to perform the target function. 

\emph{LHRS} uses LUT mapping to decompose the target function. The decomposition step is a crucial phase in this hierarchical method, as it is the starting point of the synthesis process. For this reason, it is of paramount importance to control its behavior.
The $ k $-LUT mapping technique that is used in \emph{LHRS} originates from the open source logic synthesis tool \textit{abc}~\cite{abc} and has been originally designed for the synthesis of classical circuits. 

In this work, we develop an alternative hierarchical framework: \textit{Resource-constrained Oracle Synthesis} (ROS). It embeds a new quan\-tum-aware LUT-mapper, that is specifically designed for the application into a hierarchical synthesis framework. 
Classical LUT-mappers, like the one used by \emph{LHRS}, aim at minimizing area and delay, but none of them have an immediate direct counterpart in quantum circuit synthesis. Instead, our mapper is designed to minimize metrics that make each LUT easier to be synthesized into a quantum circuit.

We show that the hierarchical flow that integrates our new mapper achieves a consistent reduction in the number of gates of the final circuits. 
We also integrate in the flow a method for quantum garbage management that has been proposed in~\cite{meuli19}. This method enables to efficiently uncompute intermediate results, giving control on the number of extra qubits (ancillae) of the circuit.

We show that our approach can effectively improve the state-of-the-art both in number of qubits and in number of operations. Rather than providing a method that generates additional Pareto optimal synthesis results, our approach systematically beats existing ones by improving the qubit count while not increasing the gate count---and vice versa---by improving the gate count while not increasing the qubit count.
We apply the ROS flow to synthesize quantum oracles, which could be applied in algorithms such as Grover's search, and compare our results with the state-of-the-art hierarchical method (\emph{LHRS}) showing improved results. 
\section{Preliminaries}\label{Pre}
\subsection{Quantum circuits}
Quantum computing processes qubits. A qubit can be in one of the ``classical'' logic states, 0 and 1, or in any superposition of these states. 
The state of a qubit $q$ can be defined by the linear combination of the classical states using two complex coefficients, $q = a_0|0\rangle + a_1|1\rangle$, with $a_0, a_1 \in \mathds{C}$ and $|a_0|^2 + |a_1|^2 = 1$. 

The Block sphere is a powerful representation of a qubit state. The two poles of the sphere represent the two classical states, while all the points of the sphere represent superposed states. On the equator of the block sphere there are all superposed states with $|a_0|^2 = |a_1|^2 = 1/2$ characterized by different angles with respect to the Z-axis. 

A 2-qubit system can be defined as: $q = a_{00}|00\rangle + a_{01}|01\rangle + a_{10}|10\rangle + a_{11}|11\rangle$, with $a_{00}, a_{01}, a_{10}, a_{11} \in \mathds{C}$ and $|a_{00}|^2 + |a_{01}|^2 + |a_{10}|^2 + |a_{11}|^2= 1$.  As a consequence, 4 complex coefficients are needed to represent a two-qubit state, while 8 complex coefficients are necessary to describe a 3-qubit system. In general, to represent the state of $n$ qubits and to simulate the quantum system behavior on a classical computer, $2^{n}$ complex coefficients are required.

While modeling a combinational functionality for the use in a quantum computation, it is possible to consider all the inputs as Boolean values---even when embedded as part of a quantum algorithm where entangled states in superposition are being applied.

The state of a qubit can be modified by applying quantum operations. All possible operations are reversible and can be represented by unitary matrices. Both single-qubit operations and 2-qubit operations are available, the latter changing the state of a qubit according to the state of a second one. 

There are different universal sets of quantum operations, targeting different technologies. In this work, we refer to the set that consists of the following operations: Controlled-NOT ($\text{CNOT}$), Hadamard ($H$) and rotations of an arbitrary angle $\theta$ over the Z-axis of the Block sphere ($R_z(\theta)$). All quantum operations can be represented by unitary matrices of dimension $2^n\times 2^n$, where $n$ is the number of qubits affected by the operations. For the selected universal set, the representative matrices are:
\[
\text{CNOT} = \begin{psmallmatrix}
1 & 0 & 0 & 0\\
0 & 1 & 0 & 0\\
0 & 0 & 0 & 1\\
0 & 0 & 1 & 0\\
\end{psmallmatrix}, \;
H = \frac{1}{\sqrt{2}}\begin{psmallmatrix}
1 & 1\\
1 & -1\\
\end{psmallmatrix}, \;
R_z(\theta) = \begin{psmallmatrix}
e^{\frac{-i\theta}{2}} & 0\\
0 & e^{\frac{i\theta}{2}}\\
\end{psmallmatrix}
\]

A quantum oracle is defined as a ``black box'' operation performing a multi-output Boolean function $f: \B^n \to \B^m$. The effect of an oracle $O$ performing the operation $f$ over two registers, one of $n$ qubits to store the inputs, $|x\rangle$, and one of $m$ qubits to store the outputs, $|y\rangle$, can be described as follows:
\[
O(|x\rangle \otimes |y\rangle) \mapsto |x\rangle \otimes |y\oplus f(x)\rangle
\]

The cost of a quantum circuit depends on the number of qubits required for the computation, and the number of operations that are performed. Automatic tools can be used to take into account technology constraints by synthesizing low cost quantum circuits.

\subsection{Rademacher-Walsh spectrum}
We call a function $f: \B^n \to \B$, where $\B = \{0, 1\}$, a Boolean
function over $n$ variables.  
A Boolean function can be represented by its truth table in the $\{1, -1\}$ encoding, which is a
bitstring $b_{2^n-1}b_{2^n-2}\dots b_0$ of size $2^n$ where
\[
  b_x = (-1)^{f(x_1, \dots, x_n)} \qquad \text{when $x = (x_1x_2 \dots x_n)_2$}
\]
The Hadamard transform matrix over $n$ variables is defined as:
\[
H_n = \begin{pmatrix}
H_{n-1} & H_{n-1}\\
H_{n-1} & -H_{n-1}\\
\end{pmatrix}, \quad
H_0 = 1
\]
Each row of the Hadamard transform matrix is equal to the truth table of the parity function between a subset of the $n$ variables. For example the last row of an $n$-variable Hadamard matrix will be the truth table of the parity function $p = x_1 \oplus x_2 \oplus \dots \oplus x_n$.

The Rademacher-Walsh spectrum $S$ of the function $f$ expressed as a truth table in the $\{1, -1\}$ encoding $F$ is defined as:
\[
S = H_nF
\]
Each coefficient of the spectrum represents the correlation with a parity function of a subset of the inputs. 
\begin{example}
Given the 3-input majority Boolean function $f(x_1, x_2, x_3) = \langle x_1 x_2 x_3 \rangle$, its truth table is:
\[
F = \begin{pmatrix}
1 &1 &1 &-1 &1 &-1 &-1 &-1\\
\end{pmatrix}
\]
The Rademacher-Walsh spectrum of $f$ is:
\[
S = H_3 F = 
\begin{pmatrix}
0 &4 &4 &0 &4 &0 &0 &-4\\
\end{pmatrix}
\]
\end{example}
We later make use of the fact that one can derive a quantum gate implementation for a Boolean function from the function's spectral coefficients~\cite{amy17, schuch03}.

\subsection{$k$-LUT mapping}
\begin{figure}[t]
\centering
  \subfloat[]{%
  \begin{tikzpicture}[font=\small]
    \begin{scope}[every node/.style={draw,circle,inner sep=0pt,minimum width=15pt,minimum height=15pt}]
      \node (1) at (0,0) {$n_1$};
      \node (2) at (1.5,0) {$n_2$};
      \node (3) at (0.5,.8) {$n_3$};
      \node (4) at (1,1.6) {$n_4$};
    \end{scope}
    \begin{scope}[every node/.style={inner sep=1pt}]
      \node (x1) at (-.25,-.8) {$x_1$};
      \node (x2) at (.25,-.8) {$x_2$};
      \node (x3) at (1.25,-.8) {$x_3$};
      \node (x4) at (1.75,-.8) {$x_4$};
      \node (f) at (1,2.5) {$f$};
     
     \node (a1) at (-.3, .5) {};
     \node (a2) at (-.3, .7) {};
     \node (b) at (1.75, .8) {$cut_1$};
     \node (c) at (2.2, -.4) {$cut_2$};
    \end{scope}

    \draw (1) -- (x1) (1) -- (x2); //(2) to[bend right=20] (x1) (2) -- (1) (3) -- (x2);
    \draw (2) -- (x3) (2) -- (4);
    \draw  (3) -- (x3) (4) -- (3);
    \draw [dashed] (x4) --(2) (3) -- (1);
    \draw (4) -- (f);
    \draw [dashed, color=blue] (a1) to[bend right=20] (b);
    \draw [dashed, color=red] (a2) to[bend right=20] (c);
  \end{tikzpicture}}
  \hspace{2cm}
  \subfloat[]{%
  \begin{tikzpicture}[font=\small]
    \begin{scope}[every node/.style={draw,circle,inner sep=0pt,minimum width=15pt,minimum height=15pt}]
      \node (1) at (0,0) {};
      \node (2) at (1.5,0) {};
    \end{scope}
     \begin{scope}[every node/.style={draw,circle,inner sep=0pt,minimum width=15pt,minimum height=15pt, color = blue}]
    \node (4) at (1,1.6) {};
    \end{scope}
    
    \begin{scope}[every node/.style={inner sep=1pt}]
      \node (x1) at (-.25,-.8) {$x_1$};
      \node (x2) at (.25,-.8) {$x_2$};
      \node (x3) at (1,-.8) {$x_3$};
      \node (x4) at (1.75,-.8) {$x_4$};
      \node (f) at (1,2.5) {$f$};
    \end{scope}

    \draw (1) -- (x1) (1) -- (x2); 
    \draw (2) -- (x3);
    \draw (4) -- (2) (4) -- (x3);
    \draw [dashed] (4) -- (1) (x4)--(2);
    \draw (4) -- (f);
  \end{tikzpicture}}
  \hspace{2cm}
  \subfloat[]{%
  \begin{tikzpicture}[font=\small]
    \begin{scope}[every node/.style={draw,circle,inner sep=0pt,minimum width=15pt,minimum height=15pt}]
      \node (1) at (0,0) {};
    \end{scope}
    
    \begin{scope}[every node/.style={draw,circle,inner sep=0pt,minimum width=15pt,minimum height=15pt, color = red}]
    \node (4) at (1,1.6) {};
    \end{scope}
    \begin{scope}[every node/.style={inner sep=1pt}]
      \node (x1) at (-.25,-.8) {$x_1$};
      \node (x2) at (.25,-.8) {$x_2$};
      \node (x3) at (1,-.8) {$x_3$};
      \node (x4) at (1.75,-.8) {$x_4$};
      \node (f) at (1,2.5) {$f$};
    \end{scope}

    \draw (1) -- (x1) (1) -- (x2) (x3)--(4); 

    \draw [dashed] (x4) --(4) (4) -- (1);
    \draw (4) -- (f);
  \end{tikzpicture}}
  \caption{(a) an AIG graph performing the function $f = (\overline{x}_1+\overline{x}_2)x_3\overline{x}_4$ with two possible 3-feasible cuts; (b) $3$-LUT network generated by $cut_1$; (c) $3$-LUT network generated by $cut_2$. }
  \label{AIGex}
\end{figure}
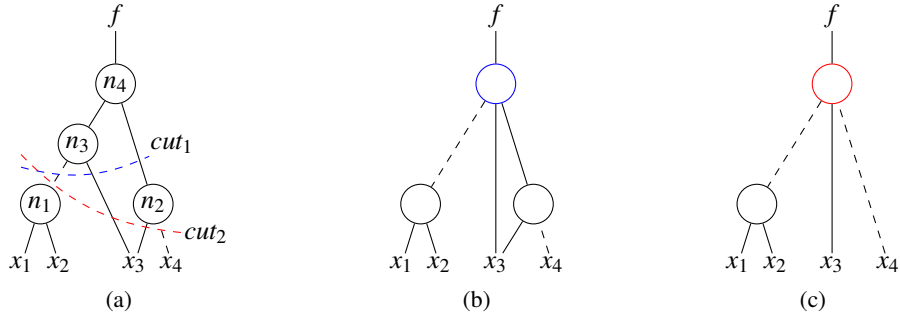

Lookup table (LUT) mapping is a decomposition method that has originally been used to map a logic design into components of FPGAs (Field Programmable Gate Array) capable of computing any Boolean function up to a given number of inputs, i.e., lookup tables.  
Later, LUT mappers found a successful application in logic synthesis and circuits optimization~\cite{mishchenko06}, as they allow to decompose large functionality into smaller functions. 
Several efficient state-of-the-art mappers are available and they are traditionally designed to minimize delay and area of the resulting circuit. 

The input of the LUT mapping is a multi-level logic network representing a Boolean function.
A multi-level logic network is represented by a graph, in which each node performs a Boolean operation and edges define data dependencies. The inputs of the function are represented by the primary inputs of the network. 
According to the characteristics of the network, we distinguish different graph representations. In And-Inverter Graphs (AIG), nodes perform the 2-input AND Boolean function, while edges can be complemented to perform inversion. A different network representation, called Xor-And-inverter Graphs (XAG) implements the 2-input XOR in addition to the 2-input AND operation. For example, a logic network representing the function $f = (\overline{x}_1+\overline{x}_2)x_3\overline{x}_4$ is shown in Fig.~\ref{AIGex}(a). All nodes of this network perform the 2-input AND operation between the node's inputs, dashed edges represent Boolean inversion, and $x_1$, $x_2$, $x_3$ and $x_4$ are the primary inputs. 

The $k$-LUT mapper decomposes the multi-level network using $k$-feasible cuts. A cut for a node $n$ is a set of leaves $l_1, \dots, l_n$ such that each path from $n$ to a primary input includes one of the leaves. A $k$-feasible cut is a cut that has at most $k$ leaves. Leaves are nodes or primary inputs of the network. In Fig.~\ref{AIGex}(a) two $3$-feasible cuts are shown for the node $n_4$. The fist cut has leaves $n_1$, $x_3$ and $n_2$, while the second cut has leaves $n_1$, $x_3$ and $x_4$. 
Fig.~\ref{AIGex}(b) and (c) show the $3$-LUT networks generated by the first and the second cut, respectively. In the first graph, the node highlighted in blue is a LUT with 3 inputs that performs the combined operations of nodes $n_4$ and $n_3$, while in the second graph, the LUT highlighted in red performs the combined operations of three nodes. 
Comparing the two networks, generated by two different cuts, it is clear how the choices made during the mapping process affect the number of nodes of the resulting LUT network and the complexity of the function performed by each LUT.

\subsection{$k$-LUT based hierarchical reversible synthesis}\label{lutexplain}
Automatic quantum circuit synthesis methods transform a high level Boolean function representation into a quantum circuit, exploiting reversible circuits as intermediate representations. 
These circuits are composed by single target gates, that are generalizations of the  multiple-controlled Toffoli gates. A single-target gate $T_f(\{x_1, \dots, x_n\}, x_k)$ is characterized by: a set of controls $x_1, \dots, x_n$, a control function $f: \B^n \rightarrow \B$ and a target $x_k$. The value of $x_k$ is complemented if the function $f(x_1, \dots, x_n)$ evaluates to one. 
Efficient methods are known for the decomposition of these reversible gates into quantum operations~\cite{Amy13, Maslov16, Meuli18best}.

Hierarchical methods for the synthesis of quantum circuits have shown the ability to synthesize large functions and enable to explore the trade-off between number of operations and number of qubits. 
Hierarchical means that the method starts from a multi-level representation of the function, i.e., a graph. 
Inputs to the function are stored on a set of existing qubits. 
Additional qubits are used to store intermedite results computed by each node of the graph.
Finally, the output results are available on some of the additional qubits.
Among the hierarchical reversible logic synthesis methods, \emph{LHRS}~\cite{Soeken18} exploits a state-of-the-art area-oriented LUT mapper, called \emph{mf} that is part of the logic synthesis framework \textit{abc}~\cite{abc} to generate a LUT network that is used as starting representation for the synthesis flow.

\begin{figure*}[t]
\centering
  \subfloat[]{
  \begin{tikzpicture}[font=\footnotesize,node distance=.38cm]

  \node[flowtool] (each) {mapping \\ into qubits \\ (Bennett)};

  \begin{scope}[every node/.style={flowbox}]
    \node[left=5cm of each] (aig) { AIG };
     \node[left=1cm of each] (klutn) { $k$-LUT \\ network };
    \node[right=1cm of each] (stg) {STG \\ network};
    \node[right=5cm of each] (qc) {quantum \\ circuit};
  \end{scope}

  \node[right=1cm of aig,flowtool] (mapping) {$k$-LUT \\ mapping};  
  \node[right=1cm of stg,flowtool] (gray) {Gray \\ synthesis};

  \begin{scope}[->]
    \draw (aig.east) to[out=0,in=180] (mapping.west);
    \draw (mapping.east) to[out=0,in=180] (klutn.west);
    \draw (klutn.east) to[out=0,in=180] (each.west);
    \draw (each.east) to[out=0,in=180] (stg.west);
    \draw (stg.east) to[out=0,in=180] (gray.west);
    \draw (gray.east) to[out=0,in=180] (qc.west);

  \end{scope}
\end{tikzpicture}
  }
  \hfill
  \subfloat[]{
  \begin{tikzpicture}[font=\footnotesize,node distance=.38cm]

  \node[flowtool, blue] (each) {SAT-based \\ memory management\\  (pebbling)};

  \begin{scope}[every node/.style={flowbox, blue}]
    \node[left=5cm of each] (aig) { XAG };
  \end{scope}
  
  \begin{scope}[every node/.style={flowbox}]
     \node[left=.8cm of each] (klutn) { $k$-LUT \\ network };
    \node[right=1cm of each] (stg) {STG \\ network};
    \node[right=5cm of each] (qc) {quantum \\ circuit};
  \end{scope}

  \node[right=.6cm of aig,flowtool, blue] (mapping) { quantum-aware \\ $k$-LUT \\ mapping};  
  \node[right=1cm of stg,flowtool] (gray) {Gray \\ synthesis};

  \begin{scope}[->]
    \draw (aig.east) to[out=0,in=180] (mapping.west);
    \draw (mapping.east) to[out=0,in=180] (klutn.west);
    \draw (klutn.east) to[out=0,in=180] (each.west);
    \draw (each.east) to[out=0,in=180] (stg.west);
    \draw (stg.east) to[out=0,in=180] (gray.west);
    \draw (gray.east) to[out=0,in=180] (qc.west);

  \end{scope}
\end{tikzpicture}
  }
  \caption{(a) state-of-the-art hierarchical synthesis framework; (b) proposed hierarchical synthesis framework.}
  \label{fig:flow}
  \cutimage
\end{figure*}
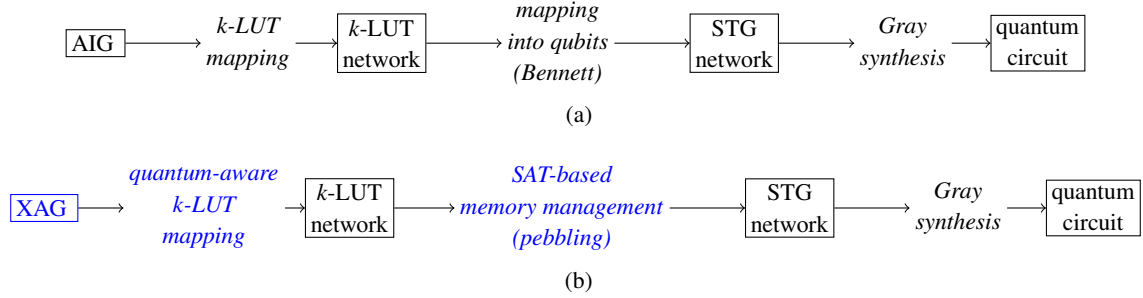

The flow of \emph{LHRS} is shown in Fig.~\ref{fig:flow}(a). 
The first step of the flow, i.e., the $k$-LUT mapping, has a large impact on the final result, as it defines: (i) the number of required qubits, and (ii) the complexity of each sub-network. Fig.~\ref{fig:example}(a) show an input network that is transformed by the mapper into the $2$-LUT network in Fig.~\ref{fig:example}(b).
In the successive step, the $k$-LUT network is transformed into a reversible circuit, a network made of single-target gates (STG network). In this reversible representation, each line corresponds to a single qubit. This step is performed by exploiting the one-to-one correspondence between nodes of the $k$-LUT network and reversible single-target gates. It transforms the network in Fig.~\ref{fig:example}(b) into one of the STG networks in Fig.~\ref{fig:example}(c) and (d).
The Gray synthesis method~\cite{amy17, schuch03} (see Fig.~\ref{fig:flow}(a)) decomposes each single-target gate with control function $f$ into a quantum circuit that consists of the following quantum operations: $\text{CNOT}$, $H$, $R_z(\theta)$. 
The method is characterized by a direct dependence between nonzero coefficients in the Hadamard-Walsh spectrum of the function $f$ and number of $\text{CNOT}$ and $R_\theta$ gates to be synthesized. This method is performed after the $k$-LUT mapping, this means that by modifying the mapping we get some control on the characteristics of the control functions in the LUT network, that are input to the Gray synthesis. 

In \emph{LHRS}, $k$-LUT mapping is performed considering metrics as delay and area, that have no immediate correlation in this application; so when $k$-LUT mapping is used in the context of quantum circuit synthesis, the classical metrics must be changed to context-related ones.
In this work we address this criticality by integrating in ROS a new quantum-aware $k$-LUT mapper that aims at minimizing the number of gates required to synthesize the quantum circuit of each LUT using the Gray synthesis method.

\subsection{Quantum memory management}
Quantum circuits are required to be garbage free, that is, any intermediate result needs to be accessible from the outputs. Otherwise, as many states can be entangled together, measurement of intermediate results may compromise the computation. 
Recently, a method for quantum memory management has been proposed that is based on solving instances of the reversible pebbling game~\cite{meuli19}. This technique can grant control on the number of qubits that are used in the clean-up process, exploiting a state-of-the-art SAT solver~\cite{Nik08}.
In~\cite{meuli19}, the authors show how the problem of uncomputing intermediate results corresponds to the reversible pebbling game.

\begin{figure}[t]
\centering
  \subfloat[]{%
  \begin{tikzpicture}[font=\small]
    \begin{scope}[every node/.style={draw,circle,inner sep=0pt,minimum width=11pt,minimum height=11pt}]
      \node (1) at (0,0) {};
      \node (2) at (0,.5) {};
      \node (3) at (0.5,0.5) {};
      \node (4) at (0.25,1) {$n_1$};
      \node (5) at (0.5,1.5) {$n_3$};
      \node (6) at (1,1.5) {$n_2$};
      \node (7) at (0.75,2) {};
      \node (8) at (1,2.5) {$n_4$};
    \end{scope}
    \begin{scope}[every node/.style={inner sep=1pt}]
      \node (x1) at (-.25,-.5) {$x_1$};
      \node (x2) at (.25,-.5) {$x_2$};
      \node (x3) at (1.25,1) {$x_3$};
      \node (f) at (1,3) {$f$};
    \end{scope}

    \draw (1) -- (x1) (1) -- (x2) (2) to[bend right=20] (x1) (2) -- (1) (3) -- (x2);
    \draw (4) -- (2) (4) -- (3) (5) -- (4) (5) to[bend left=33] (x2);
    \draw (6) to[bend left=20] (x2) (6) -- (x3) (7) -- (5) (7) -- (6) (8) -- (7) (8) -- (6);
    \draw (8) -- (f);
  \end{tikzpicture}}
  \hspace{2cm}
  \subfloat[]{%
  \begin{tikzpicture}[font=\small]
    \begin{scope}[every node/.style={draw,circle,inner sep=0pt,minimum width=11pt,minimum height=11pt}]
      \node (1) at (0,0) {$n_1$};
      \node (2) at (.5,0) {$n_2$};
      \node (3) at (.25,.5) {$n_3$};
      \node (4) at (.375,1) {$n_4$};
    \end{scope}
    \begin{scope}[every node/.style={inner sep=1pt}]
      \node (x1) at (-.25,-.5) {$x_1$};
      \node (x2) at (.25,-.5) {$x_2$};
      \node (x3) at (.75,-.5) {$x_3$};
      \node (f) at (.375,1.5) {$f$};
    \end{scope}

    \draw (1) -- (x1) (1) -- (x2) (2) -- (x2) (2) -- (x3) (3) -- (1) (3) -- (x2) (4) -- (3) (4) to[bend left=10] (2) (f) -- (4);
  \end{tikzpicture}}
  \hfill
  \subfloat[]{%
    \small
    \tikzpicture[scale=0.750000,x=1pt,y=1pt]
\filldraw[color=white] (0.000000, -6.500000) rectangle (98.000000, 84.500000);
\draw[color=black] (0.000000,78.000000) -- (98.000000,78.000000);
\draw[color=black] (0.000000,78.000000) node[left] {$x_1$};
\draw[color=black] (0.000000,65.000000) -- (98.000000,65.000000);
\draw[color=black] (0.000000,65.000000) node[left] {$x_2$};
\draw[color=black] (0.000000,52.000000) -- (98.000000,52.000000);
\draw[color=black] (0.000000,52.000000) node[left] {$x_3$};
\draw[color=black] (0.000000,39.000000) -- (98.000000,39.000000);
\draw[color=black] (0.000000,39.000000) node[left] {$0$};
\draw[color=black] (0.000000,26.000000) -- (98.000000,26.000000);
\draw[color=black] (0.000000,26.000000) node[left] {$0$};
\draw[color=black] (0.000000,13.000000) -- (98.000000,13.000000);
\draw[color=black] (0.000000,13.000000) node[left] {$0$};
\draw[color=black] (0.000000,0.000000) -- (98.000000,0.000000);
\draw[color=black] (0.000000,0.000000) node[left] {$0$};
\draw (7.000000,78.000000) -- (7.000000,39.000000);
\scope[rounded corners=2pt]
\scope
\draw[fill=white] (7.000000, 71.500000) +(-45.000000:8.485281pt and 17.677670pt) -- +(45.000000:8.485281pt and 17.677670pt) -- +(135.000000:8.485281pt and 17.677670pt) -- +(225.000000:8.485281pt and 17.677670pt) -- cycle;
\clip (7.000000, 71.500000) +(-45.000000:8.485281pt and 17.677670pt) -- +(45.000000:8.485281pt and 17.677670pt) -- +(135.000000:8.485281pt and 17.677670pt) -- +(225.000000:8.485281pt and 17.677670pt) -- cycle;
\draw (7.000000, 71.500000) node {{$n_1$}};
\endscope
\endscope
\scope
\draw[fill=white] (7.000000, 39.000000) circle(3.000000pt);
\clip (7.000000, 39.000000) circle(3.000000pt);
\draw (4.000000, 39.000000) -- (10.000000, 39.000000);
\draw (7.000000, 36.000000) -- (7.000000, 42.000000);
\endscope
\draw (21.000000,65.000000) -- (21.000000,26.000000);
\scope[rounded corners=2pt]
\scope
\draw[fill=white] (21.000000, 58.500000) +(-45.000000:8.485281pt and 17.677670pt) -- +(45.000000:8.485281pt and 17.677670pt) -- +(135.000000:8.485281pt and 17.677670pt) -- +(225.000000:8.485281pt and 17.677670pt) -- cycle;
\clip (21.000000, 58.500000) +(-45.000000:8.485281pt and 17.677670pt) -- +(45.000000:8.485281pt and 17.677670pt) -- +(135.000000:8.485281pt and 17.677670pt) -- +(225.000000:8.485281pt and 17.677670pt) -- cycle;
\draw (21.000000, 58.500000) node {{$n_2$}};
\endscope
\endscope
\scope
\draw[fill=white] (21.000000, 26.000000) circle(3.000000pt);
\clip (21.000000, 26.000000) circle(3.000000pt);
\draw (18.000000, 26.000000) -- (24.000000, 26.000000);
\draw (21.000000, 23.000000) -- (21.000000, 29.000000);
\endscope
\draw (35.000000,65.000000) -- (35.000000,13.000000);
\scope[rounded corners=2pt]
\scope
\draw[fill=white] (35.000000, 52.000000) +(-45.000000:8.485281pt and 26.870058pt) -- +(45.000000:8.485281pt and 26.870058pt) -- +(135.000000:8.485281pt and 26.870058pt) -- +(225.000000:8.485281pt and 26.870058pt) -- cycle;
\clip (35.000000, 52.000000) +(-45.000000:8.485281pt and 26.870058pt) -- +(45.000000:8.485281pt and 26.870058pt) -- +(135.000000:8.485281pt and 26.870058pt) -- +(225.000000:8.485281pt and 26.870058pt) -- cycle;
\draw (35.000000, 52.000000) node {{$n_3$}};
\endscope
\endscope
\draw[color=black,dashed] (29.000000, 52.000000) -- (41.000000, 52.000000);
\scope
\draw[fill=white] (35.000000, 13.000000) circle(3.000000pt);
\clip (35.000000, 13.000000) circle(3.000000pt);
\draw (32.000000, 13.000000) -- (38.000000, 13.000000);
\draw (35.000000, 10.000000) -- (35.000000, 16.000000);
\endscope
\draw (49.000000,26.000000) -- (49.000000,0.000000);
\scope[rounded corners=2pt]
\scope
\draw[fill=white] (49.000000, 19.500000) +(-45.000000:8.485281pt and 17.677670pt) -- +(45.000000:8.485281pt and 17.677670pt) -- +(135.000000:8.485281pt and 17.677670pt) -- +(225.000000:8.485281pt and 17.677670pt) -- cycle;
\clip (49.000000, 19.500000) +(-45.000000:8.485281pt and 17.677670pt) -- +(45.000000:8.485281pt and 17.677670pt) -- +(135.000000:8.485281pt and 17.677670pt) -- +(225.000000:8.485281pt and 17.677670pt) -- cycle;
\draw (49.000000, 19.500000) node {{$n_4$}};
\endscope
\endscope
\scope
\draw[fill=white] (49.000000, 0.000000) circle(3.000000pt);
\clip (49.000000, 0.000000) circle(3.000000pt);
\draw (46.000000, 0.000000) -- (52.000000, 0.000000);
\draw (49.000000, -3.000000) -- (49.000000, 3.000000);
\endscope
\draw (63.000000,65.000000) -- (63.000000,13.000000);
\scope[rounded corners=2pt]
\scope
\draw[fill=white] (63.000000, 52.000000) +(-45.000000:8.485281pt and 26.870058pt) -- +(45.000000:8.485281pt and 26.870058pt) -- +(135.000000:8.485281pt and 26.870058pt) -- +(225.000000:8.485281pt and 26.870058pt) -- cycle;
\clip (63.000000, 52.000000) +(-45.000000:8.485281pt and 26.870058pt) -- +(45.000000:8.485281pt and 26.870058pt) -- +(135.000000:8.485281pt and 26.870058pt) -- +(225.000000:8.485281pt and 26.870058pt) -- cycle;
\draw (63.000000, 52.000000) node {{$n_3$}};
\endscope
\endscope
\draw[color=black,dashed] (57.000000, 52.000000) -- (69.000000, 52.000000);
\scope
\draw[fill=white] (63.000000, 13.000000) circle(3.000000pt);
\clip (63.000000, 13.000000) circle(3.000000pt);
\draw (60.000000, 13.000000) -- (66.000000, 13.000000);
\draw (63.000000, 10.000000) -- (63.000000, 16.000000);
\endscope
\draw (77.000000,65.000000) -- (77.000000,26.000000);
\scope[rounded corners=2pt]
\scope
\draw[fill=white] (77.000000, 58.500000) +(-45.000000:8.485281pt and 17.677670pt) -- +(45.000000:8.485281pt and 17.677670pt) -- +(135.000000:8.485281pt and 17.677670pt) -- +(225.000000:8.485281pt and 17.677670pt) -- cycle;
\clip (77.000000, 58.500000) +(-45.000000:8.485281pt and 17.677670pt) -- +(45.000000:8.485281pt and 17.677670pt) -- +(135.000000:8.485281pt and 17.677670pt) -- +(225.000000:8.485281pt and 17.677670pt) -- cycle;
\draw (77.000000, 58.500000) node {{$n_2$}};
\endscope
\endscope
\scope
\draw[fill=white] (77.000000, 26.000000) circle(3.000000pt);
\clip (77.000000, 26.000000) circle(3.000000pt);
\draw (74.000000, 26.000000) -- (80.000000, 26.000000);
\draw (77.000000, 23.000000) -- (77.000000, 29.000000);
\endscope
\draw (91.000000,78.000000) -- (91.000000,39.000000);
\scope[rounded corners=2pt]
\scope
\draw[fill=white] (91.000000, 71.500000) +(-45.000000:8.485281pt and 17.677670pt) -- +(45.000000:8.485281pt and 17.677670pt) -- +(135.000000:8.485281pt and 17.677670pt) -- +(225.000000:8.485281pt and 17.677670pt) -- cycle;
\clip (91.000000, 71.500000) +(-45.000000:8.485281pt and 17.677670pt) -- +(45.000000:8.485281pt and 17.677670pt) -- +(135.000000:8.485281pt and 17.677670pt) -- +(225.000000:8.485281pt and 17.677670pt) -- cycle;
\draw (91.000000, 71.500000) node {{$n_1$}};
\endscope
\endscope
\scope
\draw[fill=white] (91.000000, 39.000000) circle(3.000000pt);
\clip (91.000000, 39.000000) circle(3.000000pt);
\draw (88.000000, 39.000000) -- (94.000000, 39.000000);
\draw (91.000000, 36.000000) -- (91.000000, 42.000000);
\endscope
\draw[color=black] (98.000000,78.000000) node[right] {$x_1$};
\draw[color=black] (98.000000,65.000000) node[right] {$x_2$};
\draw[color=black] (98.000000,52.000000) node[right] {$x_3$};
\draw[color=black] (98.000000,39.000000) node[right] {$0$};
\draw[color=black] (98.000000,26.000000) node[right] {$0$};
\draw[color=black] (98.000000,13.000000) node[right] {$0$};
\draw[color=black] (98.000000,0.000000) node[right] {$f$};
\endtikzpicture
  }
  \subfloat[]{%
    \small
    \tikzpicture[scale=0.750000,x=1pt,y=1pt]
\filldraw[color=white] (0.000000, -6.500000) rectangle (126.000000, 71.500000);
\draw[color=black] (0.000000,65.000000) -- (126.000000,65.000000);
\draw[color=black] (0.000000,65.000000) node[left] {$x_1$};
\draw[color=black] (0.000000,52.000000) -- (126.000000,52.000000);
\draw[color=black] (0.000000,52.000000) node[left] {$x_2$};
\draw[color=black] (0.000000,39.000000) -- (126.000000,39.000000);
\draw[color=black] (0.000000,39.000000) node[left] {$x_3$};
\draw[color=black] (0.000000,26.000000) -- (126.000000,26.000000);
\draw[color=black] (0.000000,26.000000) node[left] {$0$};
\draw[color=black] (0.000000,13.000000) -- (126.000000,13.000000);
\draw[color=black] (0.000000,13.000000) node[left] {$0$};
\draw[color=black] (0.000000,0.000000) -- (126.000000,0.000000);
\draw[color=black] (0.000000,0.000000) node[left] {$0$};
\draw (7.000000,65.000000) -- (7.000000,26.000000);
\scope[rounded corners=2pt]
\scope
\draw[fill=white] (7.000000, 58.500000) +(-45.000000:8.485281pt and 17.677670pt) -- +(45.000000:8.485281pt and 17.677670pt) -- +(135.000000:8.485281pt and 17.677670pt) -- +(225.000000:8.485281pt and 17.677670pt) -- cycle;
\clip (7.000000, 58.500000) +(-45.000000:8.485281pt and 17.677670pt) -- +(45.000000:8.485281pt and 17.677670pt) -- +(135.000000:8.485281pt and 17.677670pt) -- +(225.000000:8.485281pt and 17.677670pt) -- cycle;
\draw (7.000000, 58.500000) node {{$n_1$}};
\endscope
\endscope
\scope
\draw[fill=white] (7.000000, 26.000000) circle(3.000000pt);
\clip (7.000000, 26.000000) circle(3.000000pt);
\draw (4.000000, 26.000000) -- (10.000000, 26.000000);
\draw (7.000000, 23.000000) -- (7.000000, 29.000000);
\endscope
\draw (21.000000,52.000000) -- (21.000000,13.000000);
\scope[rounded corners=2pt]
\scope
\draw[fill=white] (21.000000, 39.000000) +(-45.000000:8.485281pt and 26.870058pt) -- +(45.000000:8.485281pt and 26.870058pt) -- +(135.000000:8.485281pt and 26.870058pt) -- +(225.000000:8.485281pt and 26.870058pt) -- cycle;
\clip (21.000000, 39.000000) +(-45.000000:8.485281pt and 26.870058pt) -- +(45.000000:8.485281pt and 26.870058pt) -- +(135.000000:8.485281pt and 26.870058pt) -- +(225.000000:8.485281pt and 26.870058pt) -- cycle;
\draw (21.000000, 39.000000) node {{$n_3$}};
\endscope
\endscope
\draw[color=black,dashed] (15.000000, 39.000000) -- (27.000000, 39.000000);
\scope
\draw[fill=white] (21.000000, 13.000000) circle(3.000000pt);
\clip (21.000000, 13.000000) circle(3.000000pt);
\draw (18.000000, 13.000000) -- (24.000000, 13.000000);
\draw (21.000000, 10.000000) -- (21.000000, 16.000000);
\endscope
\draw (35.000000,65.000000) -- (35.000000,26.000000);
\scope[rounded corners=2pt]
\scope
\draw[fill=white] (35.000000, 58.500000) +(-45.000000:8.485281pt and 17.677670pt) -- +(45.000000:8.485281pt and 17.677670pt) -- +(135.000000:8.485281pt and 17.677670pt) -- +(225.000000:8.485281pt and 17.677670pt) -- cycle;
\clip (35.000000, 58.500000) +(-45.000000:8.485281pt and 17.677670pt) -- +(45.000000:8.485281pt and 17.677670pt) -- +(135.000000:8.485281pt and 17.677670pt) -- +(225.000000:8.485281pt and 17.677670pt) -- cycle;
\draw (35.000000, 58.500000) node {{$n_1$}};
\endscope
\endscope
\scope
\draw[fill=white] (35.000000, 26.000000) circle(3.000000pt);
\clip (35.000000, 26.000000) circle(3.000000pt);
\draw (32.000000, 26.000000) -- (38.000000, 26.000000);
\draw (35.000000, 23.000000) -- (35.000000, 29.000000);
\endscope
\draw (49.000000,52.000000) -- (49.000000,26.000000);
\scope[rounded corners=2pt]
\scope
\draw[fill=white] (49.000000, 45.500000) +(-45.000000:8.485281pt and 17.677670pt) -- +(45.000000:8.485281pt and 17.677670pt) -- +(135.000000:8.485281pt and 17.677670pt) -- +(225.000000:8.485281pt and 17.677670pt) -- cycle;
\clip (49.000000, 45.500000) +(-45.000000:8.485281pt and 17.677670pt) -- +(45.000000:8.485281pt and 17.677670pt) -- +(135.000000:8.485281pt and 17.677670pt) -- +(225.000000:8.485281pt and 17.677670pt) -- cycle;
\draw (49.000000, 45.500000) node {{$n_2$}};
\endscope
\endscope
\scope
\draw[fill=white] (49.000000, 26.000000) circle(3.000000pt);
\clip (49.000000, 26.000000) circle(3.000000pt);
\draw (46.000000, 26.000000) -- (52.000000, 26.000000);
\draw (49.000000, 23.000000) -- (49.000000, 29.000000);
\endscope
\draw (63.000000,26.000000) -- (63.000000,0.000000);
\scope[rounded corners=2pt]
\scope
\draw[fill=white] (63.000000, 19.500000) +(-45.000000:8.485281pt and 17.677670pt) -- +(45.000000:8.485281pt and 17.677670pt) -- +(135.000000:8.485281pt and 17.677670pt) -- +(225.000000:8.485281pt and 17.677670pt) -- cycle;
\clip (63.000000, 19.500000) +(-45.000000:8.485281pt and 17.677670pt) -- +(45.000000:8.485281pt and 17.677670pt) -- +(135.000000:8.485281pt and 17.677670pt) -- +(225.000000:8.485281pt and 17.677670pt) -- cycle;
\draw (63.000000, 19.500000) node {{$n_4$}};
\endscope
\endscope
\scope
\draw[fill=white] (63.000000, 0.000000) circle(3.000000pt);
\clip (63.000000, 0.000000) circle(3.000000pt);
\draw (60.000000, 0.000000) -- (66.000000, 0.000000);
\draw (63.000000, -3.000000) -- (63.000000, 3.000000);
\endscope
\draw (77.000000,52.000000) -- (77.000000,26.000000);
\scope[rounded corners=2pt]
\scope
\draw[fill=white] (77.000000, 45.500000) +(-45.000000:8.485281pt and 17.677670pt) -- +(45.000000:8.485281pt and 17.677670pt) -- +(135.000000:8.485281pt and 17.677670pt) -- +(225.000000:8.485281pt and 17.677670pt) -- cycle;
\clip (77.000000, 45.500000) +(-45.000000:8.485281pt and 17.677670pt) -- +(45.000000:8.485281pt and 17.677670pt) -- +(135.000000:8.485281pt and 17.677670pt) -- +(225.000000:8.485281pt and 17.677670pt) -- cycle;
\draw (77.000000, 45.500000) node {{$n_2$}};
\endscope
\endscope
\scope
\draw[fill=white] (77.000000, 26.000000) circle(3.000000pt);
\clip (77.000000, 26.000000) circle(3.000000pt);
\draw (74.000000, 26.000000) -- (80.000000, 26.000000);
\draw (77.000000, 23.000000) -- (77.000000, 29.000000);
\endscope
\draw (91.000000,65.000000) -- (91.000000,26.000000);
\scope[rounded corners=2pt]
\scope
\draw[fill=white] (91.000000, 58.500000) +(-45.000000:8.485281pt and 17.677670pt) -- +(45.000000:8.485281pt and 17.677670pt) -- +(135.000000:8.485281pt and 17.677670pt) -- +(225.000000:8.485281pt and 17.677670pt) -- cycle;
\clip (91.000000, 58.500000) +(-45.000000:8.485281pt and 17.677670pt) -- +(45.000000:8.485281pt and 17.677670pt) -- +(135.000000:8.485281pt and 17.677670pt) -- +(225.000000:8.485281pt and 17.677670pt) -- cycle;
\draw (91.000000, 58.500000) node {{$n_1$}};
\endscope
\endscope
\scope
\draw[fill=white] (91.000000, 26.000000) circle(3.000000pt);
\clip (91.000000, 26.000000) circle(3.000000pt);
\draw (88.000000, 26.000000) -- (94.000000, 26.000000);
\draw (91.000000, 23.000000) -- (91.000000, 29.000000);
\endscope
\draw (105.000000,52.000000) -- (105.000000,13.000000);
\scope[rounded corners=2pt]
\scope
\draw[fill=white] (105.000000, 39.000000) +(-45.000000:8.485281pt and 26.870058pt) -- +(45.000000:8.485281pt and 26.870058pt) -- +(135.000000:8.485281pt and 26.870058pt) -- +(225.000000:8.485281pt and 26.870058pt) -- cycle;
\clip (105.000000, 39.000000) +(-45.000000:8.485281pt and 26.870058pt) -- +(45.000000:8.485281pt and 26.870058pt) -- +(135.000000:8.485281pt and 26.870058pt) -- +(225.000000:8.485281pt and 26.870058pt) -- cycle;
\draw (105.000000, 39.000000) node {{$n_3$}};
\endscope
\endscope
\draw[color=black,dashed] (99.000000, 39.000000) -- (111.000000, 39.000000);
\scope
\draw[fill=white] (105.000000, 13.000000) circle(3.000000pt);
\clip (105.000000, 13.000000) circle(3.000000pt);
\draw (102.000000, 13.000000) -- (108.000000, 13.000000);
\draw (105.000000, 10.000000) -- (105.000000, 16.000000);
\endscope
\draw (119.000000,65.000000) -- (119.000000,26.000000);
\scope[rounded corners=2pt]
\scope
\draw[fill=white] (119.000000, 58.500000) +(-45.000000:8.485281pt and 17.677670pt) -- +(45.000000:8.485281pt and 17.677670pt) -- +(135.000000:8.485281pt and 17.677670pt) -- +(225.000000:8.485281pt and 17.677670pt) -- cycle;
\clip (119.000000, 58.500000) +(-45.000000:8.485281pt and 17.677670pt) -- +(45.000000:8.485281pt and 17.677670pt) -- +(135.000000:8.485281pt and 17.677670pt) -- +(225.000000:8.485281pt and 17.677670pt) -- cycle;
\draw (119.000000, 58.500000) node {{$n_1$}};
\endscope
\endscope
\scope
\draw[fill=white] (119.000000, 26.000000) circle(3.000000pt);
\clip (119.000000, 26.000000) circle(3.000000pt);
\draw (116.000000, 26.000000) -- (122.000000, 26.000000);
\draw (119.000000, 23.000000) -- (119.000000, 29.000000);
\endscope
\draw[color=black] (126.000000,65.000000) node[right] {$x_1$};
\draw[color=black] (126.000000,52.000000) node[right] {$x_2$};
\draw[color=black] (126.000000,39.000000) node[right] {$x_3$};
\draw[color=black] (126.000000,26.000000) node[right] {$0$};
\draw[color=black] (126.000000,13.000000) node[right] {$0$};
\draw[color=black] (126.000000,0.000000) node[right] {$f$};
\endtikzpicture
  }
  \caption{(a) a three-input multi-level logic network performing the Booelan function $f$; (b) an example of a $2$-LUT network for $f$; (c) the reversible circuit for the $2$-LUT network obtained using the Bennett clean-up strategy; (d) the reversible circuit for the $2$-LUT network obtained using the quantum garbage management clean-up strategy~\cite{meuli19}.}
  \label{fig:example}
  \cutimage
\end{figure}
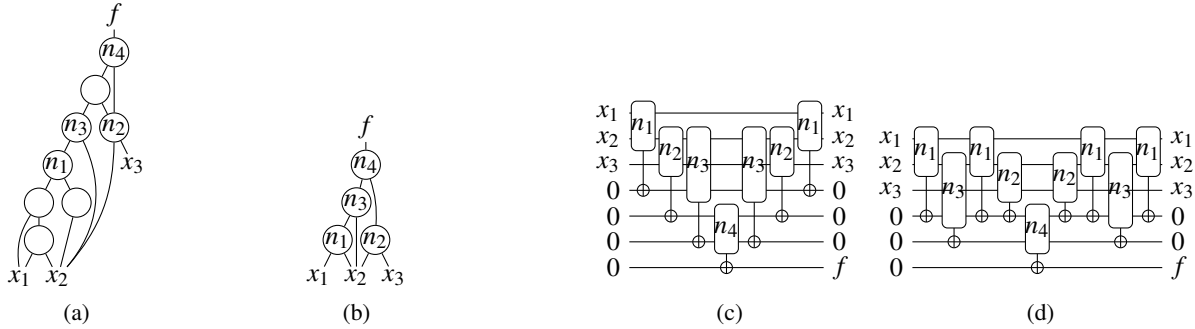

Consider Fig.~\ref{fig:example}(c), here the $k$-LUT network is mapped into qubits, and each node is transformed in its corresponding reversible gate. The intermediate result is stored on a qubit that was initialized to $|0\rangle$. After the result is computed, all the intermediate values $n_1, n_2$, and $n_3$ must be uncomputed. This is done by performing the same operation twice. The order in which nodes are computed and uncomputed is a clean-up strategy, in Fig.~\ref{fig:example}(c) the strategy used is called Bennett strategy~\cite{Bennett1989}. 
The SAT-based method described in~\cite{meuli19} is capable of finding clean-up strategies that reduce the number of required qubits by computing and uncomputing the same reversible operation more than once. An example is shown in Fig.~\ref{fig:example}(d).

\section{Resource-constrained Oracle synthesis}
Even if quantum computing is promising to beat its classical counterpart in many applications, quantum device technology is still developing and is fairly new with respect to standard CMOS technology. With our tool we aim at providing the designer with the capability to tune the synthesis with respect to the available hardware.

We propose ROS, a hierarchical synthesis framework built to leverage the quantum circuit cost, both in terms of number of qubits \emph{and} number of gates. ROS's synthesis flow is shown in Fig.~\ref{fig:flow}(b). It introduces two main contributions with respect to the state-of-the-art flow (see Fig.~\ref{fig:flow}(a)). 

\begin{itemize}
\item{
First, it embeds a new $k$-LUT mapper that is used to decompose the initial functionality into LUTs, in such a way to minimize the cost of each LUT. Such cost is defined according to the complexity of the LUT function to be synthesized by the Gray algorithm. If we analyze the result of this mapper against the state-of-the-art \emph{mf} mapper, we get in general more LUTs, each one corresponding to fewer gates. 
}
\item{
Second, it exploits the quantum garbage management technique presented in~\cite{meuli19} to control the number of qubits. 
}
\end{itemize}

\begin{figure}[t]
\centering
\begin{tikzpicture}[font=\small]

  \draw[->] (0,0) -- node[rotate=90,above] {\#qubits} ++(up:3.2cm);
  \draw[->] (0,0) -- node[below] {\#gates} ++(right:3.2cm);

  \draw[dashed] (0,3) -- (3,0);

  \draw[fill=white] (0.5,2.5) circle (2pt) node[right] {$S/B$};
  \draw[fill=white] (1.5,1.5) circle (2pt) node[right] {$M/B$};
  \draw[fill=white] (2.5,0.5) circle (2pt) node[right] {$M/P$};

  \draw[fill=white] (0.8,1.5) circle (2pt) node[above] {$S/P$};
  \draw[fill=white] (1.5,0.8) circle (2pt) node[right] {$S/P$};

  \draw[->,shorten <=3pt, shorten >=3pt] (1.5,1.5) -- (0.8,1.5);
  \draw[->,shorten <=3pt, shorten >=3pt] (1.5,1.5) -- (1.5,0.8);

  \node[right,align=left] at (3,2.7) {$S$: spectral mapper (new) \\ $M$: mf mapper (state-of-the-art)};
  \node[right,align=left] at (3,1.7) {$P$: memory management technique \\(state-of-the-art) \\ $B$: Bennett (state-of-the-art)};
\end{tikzpicture}
\caption{qualitative description of ROS's capability}
\label{fig:qual}
\end{figure}
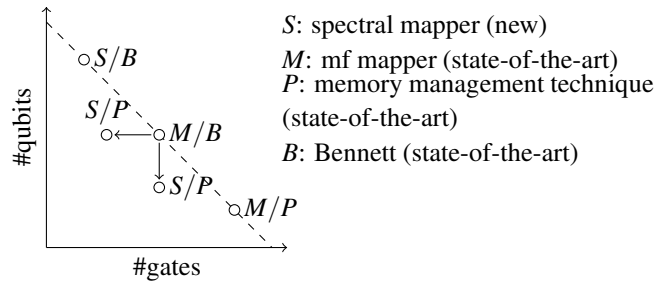

We claim that by using those two techniques together with the Gray synthesis method, we can beat the state-of-the-art results both in number of qubits and number of gates.
Fig.~\ref{fig:qual} shows a qualitative description of the performance advantages we expect to obtain using ROS. In the plot, the state-of-the-art result is the one marked as $M/B$ (corresponding to the \emph{LHRS} synthesis framework). If we only apply the memory management technique in~\cite{meuli19}, but not the new mapper, we will obtain a circuit with fewer qubits and more gates, that correspond in the figure to $M/P$. If instead we embed into \emph{LHRS} our quantum aware $k$-LUT mapper, but no quantum memory management, we can obtain a circuit with few gates but more qubits: $S/B$. Only by combining both techniques ($S/P$), we can beat the state-of-the-art tool in both qubits and gates.  In fact, we can tune the approach to only improve qubit count by not increase gate count, or vice versa. This qualitative description is supported by the results in Section~\ref{results}.

\section{Quantum-aware $k$-LUT mapping}
A main contribution of this work is to develop a $k$-LUT mapper designed to reduce the resources needed to synthesize the quantum circuit of each LUT. 
As explained in Section~\ref{lutexplain}, LUT mapping is used to decompose a logic design that is too large to be synthesized into a quantum circuit by the existing methods.  
In this section, we describe how to perform this decomposition to facilitate the successive synthesis steps.

\subsection{Cut enumeration and costing}
An Xor-And-inverter Graph, or XAG, is input to the $k$-LUT mapping process. This is a graph where each node performs the XOR or the AND operation, and where edges can be complemented to perform inversion. The choice of this particular data structure is not accidental, but reflects the fact that it is relatively cheap to perform the XOR operation in fault tolerant quantum circuits. Only one CNOT gate is needed to perform the XOR between two qubits, and in general $m-1$ CNOT gates are needed to perform the CNOT of an $m$-input XOR gate. If the result should be stored on a free ancilla line, $m$ CNOT gates are needed in total.

Once the input is defined, we first perform cut enumeration. This step consists of enumerating all possible cuts for each node of the input graph, traversing the graph from the bottom to the top. Only the best cuts are stored for each node, a technique called ``priority cuts'' to reduce the memory requirement of cut enumeration~\cite{mishchenko07}. During cut enumeration each node is assigned to a set of $p$ cuts that are ordered following a user defined cost criteria. In our case, as we aim at integrating the mapper into a quantum synthesis framework, we use as cost function the number of nonzero spectral coefficients of the function performed by the selected cut. As pointed out in the preliminaries, the Gray synthesis method generates smaller quantum circuits when the input function presents a spectrum with many zeros.
At the end of the cut enumeration step, each node of the XAG is assigned with an ordered set of $p$ cuts, with the order criteria defined by the spectrum of the cut's function. 

\subsection{XOR-block matching}
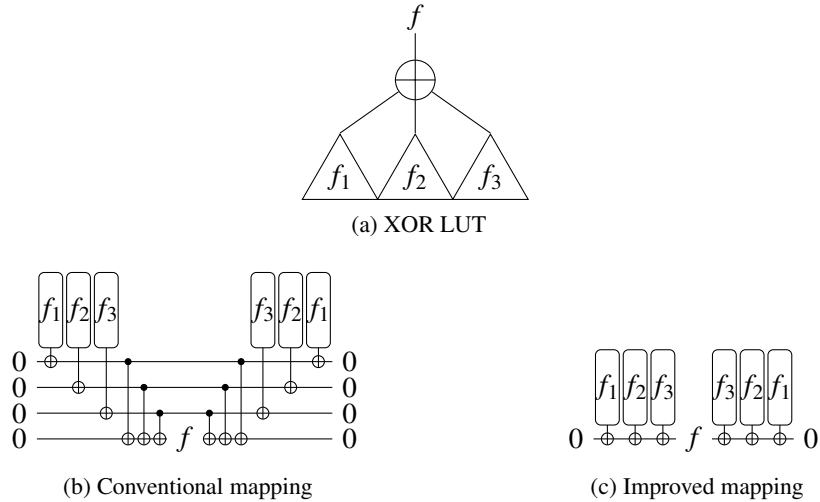
\begin{figure}[t]
  \centering
  \subfloat[XOR LUT]{\begin{tikzpicture}
  \node[draw,regular polygon,regular polygon sides=3,inner sep=1pt] (f1) {$f_1$};
  \node[draw,regular polygon,regular polygon sides=3,inner sep=1pt] (f2) at (1,0) {$f_2$};
  \node[draw,regular polygon,regular polygon sides=3,inner sep=1pt] (f3) at (2,0 ){$f_3$};

  \node[draw,circle,minimum size=15pt] (xor) at (1,1.3) {};
  \draw (xor) -- (f1.north);
  \draw (xor) -- (f2.north);
  \draw (xor) -- (f3.north);
  \draw (xor.north) -- (xor.south) (xor.west) -- (xor.east);
  \draw (xor.north) -- ++(up:10pt) node[above,inner sep=1pt] {$f$};
\end{tikzpicture}

}

  \subfloat[Conventional mapping]{\begin{tikzpicture}[scale=0.750000,x=1pt,y=1pt]
\filldraw[color=white] (0.000000, -6.500000) rectangle (149.000000, 84.500000);
\draw[color=white] (0.000000,78.000000) -- (149.000000,78.000000);
\draw[color=white] (0.000000,65.000000) -- (149.000000,65.000000);
\draw[color=white] (0.000000,52.000000) -- (149.000000,52.000000);
\draw[color=black] (0.000000,39.000000) -- (149.000000,39.000000);
\draw[color=black] (0.000000,39.000000) node[left] {$0$};
\draw[color=black] (0.000000,26.000000) -- (149.000000,26.000000);
\draw[color=black] (0.000000,26.000000) node[left] {$0$};
\draw[color=black] (0.000000,13.000000) -- (149.000000,13.000000);
\draw[color=black] (0.000000,13.000000) node[left] {$0$};
\draw[color=black] (0.000000,0.000000) -- (149.000000,0.000000);
\draw[color=black] (0.000000,0.000000) node[left] {$0$};
\draw (7.000000,78.000000) -- (7.000000,39.000000);
\begin{scope}[rounded corners=2pt]
\begin{scope}
\draw[fill=white] (7.000000, 65.000000) +(-45.000000:8.485281pt and 26.870058pt) -- +(45.000000:8.485281pt and 26.870058pt) -- +(135.000000:8.485281pt and 26.870058pt) -- +(225.000000:8.485281pt and 26.870058pt) -- cycle;
\clip (7.000000, 65.000000) +(-45.000000:8.485281pt and 26.870058pt) -- +(45.000000:8.485281pt and 26.870058pt) -- +(135.000000:8.485281pt and 26.870058pt) -- +(225.000000:8.485281pt and 26.870058pt) -- cycle;
\draw (7.000000, 65.000000) node {{$f_1$}};
\end{scope}
\end{scope}
\begin{scope}
\draw[fill=white] (7.000000, 39.000000) circle(3.000000pt);
\clip (7.000000, 39.000000) circle(3.000000pt);
\draw (4.000000, 39.000000) -- (10.000000, 39.000000);
\draw (7.000000, 36.000000) -- (7.000000, 42.000000);
\end{scope}
\draw (21.000000,78.000000) -- (21.000000,26.000000);
\begin{scope}[rounded corners=2pt]
\begin{scope}
\draw[fill=white] (21.000000, 65.000000) +(-45.000000:8.485281pt and 26.870058pt) -- +(45.000000:8.485281pt and 26.870058pt) -- +(135.000000:8.485281pt and 26.870058pt) -- +(225.000000:8.485281pt and 26.870058pt) -- cycle;
\clip (21.000000, 65.000000) +(-45.000000:8.485281pt and 26.870058pt) -- +(45.000000:8.485281pt and 26.870058pt) -- +(135.000000:8.485281pt and 26.870058pt) -- +(225.000000:8.485281pt and 26.870058pt) -- cycle;
\draw (21.000000, 65.000000) node {{$f_2$}};
\end{scope}
\end{scope}
\begin{scope}
\draw[fill=white] (21.000000, 26.000000) circle(3.000000pt);
\clip (21.000000, 26.000000) circle(3.000000pt);
\draw (18.000000, 26.000000) -- (24.000000, 26.000000);
\draw (21.000000, 23.000000) -- (21.000000, 29.000000);
\end{scope}
\draw (35.000000,78.000000) -- (35.000000,13.000000);
\begin{scope}[rounded corners=2pt]
\begin{scope}
\draw[fill=white] (35.000000, 65.000000) +(-45.000000:8.485281pt and 26.870058pt) -- +(45.000000:8.485281pt and 26.870058pt) -- +(135.000000:8.485281pt and 26.870058pt) -- +(225.000000:8.485281pt and 26.870058pt) -- cycle;
\clip (35.000000, 65.000000) +(-45.000000:8.485281pt and 26.870058pt) -- +(45.000000:8.485281pt and 26.870058pt) -- +(135.000000:8.485281pt and 26.870058pt) -- +(225.000000:8.485281pt and 26.870058pt) -- cycle;
\draw (35.000000, 65.000000) node {{$f_3$}};
\end{scope}
\end{scope}
\begin{scope}
\draw[fill=white] (35.000000, 13.000000) circle(3.000000pt);
\clip (35.000000, 13.000000) circle(3.000000pt);
\draw (32.000000, 13.000000) -- (38.000000, 13.000000);
\draw (35.000000, 10.000000) -- (35.000000, 16.000000);
\end{scope}
\draw (46.000000,39.000000) -- (46.000000,0.000000);
\filldraw (46.000000, 39.000000) circle(1.500000pt);
\begin{scope}
\draw[fill=white] (46.000000, 0.000000) circle(3.000000pt);
\clip (46.000000, 0.000000) circle(3.000000pt);
\draw (43.000000, 0.000000) -- (49.000000, 0.000000);
\draw (46.000000, -3.000000) -- (46.000000, 3.000000);
\end{scope}
\draw (54.000000,26.000000) -- (54.000000,0.000000);
\filldraw (54.000000, 26.000000) circle(1.500000pt);
\begin{scope}
\draw[fill=white] (54.000000, 0.000000) circle(3.000000pt);
\clip (54.000000, 0.000000) circle(3.000000pt);
\draw (51.000000, 0.000000) -- (57.000000, 0.000000);
\draw (54.000000, -3.000000) -- (54.000000, 3.000000);
\end{scope}
\draw (62.000000,13.000000) -- (62.000000,0.000000);
\filldraw (62.000000, 13.000000) circle(1.500000pt);
\begin{scope}
\draw[fill=white] (62.000000, 0.000000) circle(3.000000pt);
\clip (62.000000, 0.000000) circle(3.000000pt);
\draw (59.000000, 0.000000) -- (65.000000, 0.000000);
\draw (62.000000, -3.000000) -- (62.000000, 3.000000);
\end{scope}
\draw[color=black] (74.500000, 0.000000) node [fill=white] {$f$};
\draw (87.000000,13.000000) -- (87.000000,0.000000);
\filldraw (87.000000, 13.000000) circle(1.500000pt);
\begin{scope}
\draw[fill=white] (87.000000, 0.000000) circle(3.000000pt);
\clip (87.000000, 0.000000) circle(3.000000pt);
\draw (84.000000, 0.000000) -- (90.000000, 0.000000);
\draw (87.000000, -3.000000) -- (87.000000, 3.000000);
\end{scope}
\draw (95.000000,26.000000) -- (95.000000,0.000000);
\filldraw (95.000000, 26.000000) circle(1.500000pt);
\begin{scope}
\draw[fill=white] (95.000000, 0.000000) circle(3.000000pt);
\clip (95.000000, 0.000000) circle(3.000000pt);
\draw (92.000000, 0.000000) -- (98.000000, 0.000000);
\draw (95.000000, -3.000000) -- (95.000000, 3.000000);
\end{scope}
\draw (103.000000,39.000000) -- (103.000000,0.000000);
\filldraw (103.000000, 39.000000) circle(1.500000pt);
\begin{scope}
\draw[fill=white] (103.000000, 0.000000) circle(3.000000pt);
\clip (103.000000, 0.000000) circle(3.000000pt);
\draw (100.000000, 0.000000) -- (106.000000, 0.000000);
\draw (103.000000, -3.000000) -- (103.000000, 3.000000);
\end{scope}
\draw (114.000000,78.000000) -- (114.000000,13.000000);
\begin{scope}[rounded corners=2pt]
\begin{scope}
\draw[fill=white] (114.000000, 65.000000) +(-45.000000:8.485281pt and 26.870058pt) -- +(45.000000:8.485281pt and 26.870058pt) -- +(135.000000:8.485281pt and 26.870058pt) -- +(225.000000:8.485281pt and 26.870058pt) -- cycle;
\clip (114.000000, 65.000000) +(-45.000000:8.485281pt and 26.870058pt) -- +(45.000000:8.485281pt and 26.870058pt) -- +(135.000000:8.485281pt and 26.870058pt) -- +(225.000000:8.485281pt and 26.870058pt) -- cycle;
\draw (114.000000, 65.000000) node {{$f_3$}};
\end{scope}
\end{scope}
\begin{scope}
\draw[fill=white] (114.000000, 13.000000) circle(3.000000pt);
\clip (114.000000, 13.000000) circle(3.000000pt);
\draw (111.000000, 13.000000) -- (117.000000, 13.000000);
\draw (114.000000, 10.000000) -- (114.000000, 16.000000);
\end{scope}
\draw (128.000000,78.000000) -- (128.000000,26.000000);
\begin{scope}[rounded corners=2pt]
\begin{scope}
\draw[fill=white] (128.000000, 65.000000) +(-45.000000:8.485281pt and 26.870058pt) -- +(45.000000:8.485281pt and 26.870058pt) -- +(135.000000:8.485281pt and 26.870058pt) -- +(225.000000:8.485281pt and 26.870058pt) -- cycle;
\clip (128.000000, 65.000000) +(-45.000000:8.485281pt and 26.870058pt) -- +(45.000000:8.485281pt and 26.870058pt) -- +(135.000000:8.485281pt and 26.870058pt) -- +(225.000000:8.485281pt and 26.870058pt) -- cycle;
\draw (128.000000, 65.000000) node {{$f_2$}};
\end{scope}
\end{scope}
\begin{scope}
\draw[fill=white] (128.000000, 26.000000) circle(3.000000pt);
\clip (128.000000, 26.000000) circle(3.000000pt);
\draw (125.000000, 26.000000) -- (131.000000, 26.000000);
\draw (128.000000, 23.000000) -- (128.000000, 29.000000);
\end{scope}
\draw (142.000000,78.000000) -- (142.000000,39.000000);
\begin{scope}[rounded corners=2pt]
\begin{scope}
\draw[fill=white] (142.000000, 65.000000) +(-45.000000:8.485281pt and 26.870058pt) -- +(45.000000:8.485281pt and 26.870058pt) -- +(135.000000:8.485281pt and 26.870058pt) -- +(225.000000:8.485281pt and 26.870058pt) -- cycle;
\clip (142.000000, 65.000000) +(-45.000000:8.485281pt and 26.870058pt) -- +(45.000000:8.485281pt and 26.870058pt) -- +(135.000000:8.485281pt and 26.870058pt) -- +(225.000000:8.485281pt and 26.870058pt) -- cycle;
\draw (142.000000, 65.000000) node {{$f_1$}};
\end{scope}
\end{scope}
\begin{scope}
\draw[fill=white] (142.000000, 39.000000) circle(3.000000pt);
\clip (142.000000, 39.000000) circle(3.000000pt);
\draw (139.000000, 39.000000) -- (145.000000, 39.000000);
\draw (142.000000, 36.000000) -- (142.000000, 42.000000);
\end{scope}
\draw[color=black] (149.000000,39.000000) node[right] {$0$};
\draw[color=black] (149.000000,26.000000) node[right] {$0$};
\draw[color=black] (149.000000,13.000000) node[right] {$0$};
\draw[color=black] (149.000000,0.000000) node[right] {$0$};
\end{tikzpicture}}
  \hfil
  \subfloat[Improved mapping]{\begin{tikzpicture}[scale=0.750000,x=1pt,y=1pt]
\filldraw[color=white] (0.000000, -6.500000) rectangle (101.000000, 45.500000);
\draw[color=white] (0.000000,39.000000) -- (101.000000,39.000000);
\draw[color=white] (0.000000,26.000000) -- (101.000000,26.000000);
\draw[color=white] (0.000000,13.000000) -- (101.000000,13.000000);
\draw[color=black] (0.000000,0.000000) -- (101.000000,0.000000);
\draw[color=black] (0.000000,0.000000) node[left] {$0$};
\draw (7.000000,39.000000) -- (7.000000,0.000000);
\begin{scope}[rounded corners=2pt]
\begin{scope}
\draw[fill=white] (7.000000, 26.000000) +(-45.000000:8.485281pt and 26.870058pt) -- +(45.000000:8.485281pt and 26.870058pt) -- +(135.000000:8.485281pt and 26.870058pt) -- +(225.000000:8.485281pt and 26.870058pt) -- cycle;
\clip (7.000000, 26.000000) +(-45.000000:8.485281pt and 26.870058pt) -- +(45.000000:8.485281pt and 26.870058pt) -- +(135.000000:8.485281pt and 26.870058pt) -- +(225.000000:8.485281pt and 26.870058pt) -- cycle;
\draw (7.000000, 26.000000) node {{$f_1$}};
\end{scope}
\end{scope}
\begin{scope}
\draw[fill=white] (7.000000, 0.000000) circle(3.000000pt);
\clip (7.000000, 0.000000) circle(3.000000pt);
\draw (4.000000, 0.000000) -- (10.000000, 0.000000);
\draw (7.000000, -3.000000) -- (7.000000, 3.000000);
\end{scope}
\draw (21.000000,39.000000) -- (21.000000,0.000000);
\begin{scope}[rounded corners=2pt]
\begin{scope}
\draw[fill=white] (21.000000, 26.000000) +(-45.000000:8.485281pt and 26.870058pt) -- +(45.000000:8.485281pt and 26.870058pt) -- +(135.000000:8.485281pt and 26.870058pt) -- +(225.000000:8.485281pt and 26.870058pt) -- cycle;
\clip (21.000000, 26.000000) +(-45.000000:8.485281pt and 26.870058pt) -- +(45.000000:8.485281pt and 26.870058pt) -- +(135.000000:8.485281pt and 26.870058pt) -- +(225.000000:8.485281pt and 26.870058pt) -- cycle;
\draw (21.000000, 26.000000) node {{$f_2$}};
\end{scope}
\end{scope}
\begin{scope}
\draw[fill=white] (21.000000, 0.000000) circle(3.000000pt);
\clip (21.000000, 0.000000) circle(3.000000pt);
\draw (18.000000, 0.000000) -- (24.000000, 0.000000);
\draw (21.000000, -3.000000) -- (21.000000, 3.000000);
\end{scope}
\draw (35.000000,39.000000) -- (35.000000,0.000000);
\begin{scope}[rounded corners=2pt]
\begin{scope}
\draw[fill=white] (35.000000, 26.000000) +(-45.000000:8.485281pt and 26.870058pt) -- +(45.000000:8.485281pt and 26.870058pt) -- +(135.000000:8.485281pt and 26.870058pt) -- +(225.000000:8.485281pt and 26.870058pt) -- cycle;
\clip (35.000000, 26.000000) +(-45.000000:8.485281pt and 26.870058pt) -- +(45.000000:8.485281pt and 26.870058pt) -- +(135.000000:8.485281pt and 26.870058pt) -- +(225.000000:8.485281pt and 26.870058pt) -- cycle;
\draw (35.000000, 26.000000) node {{$f_3$}};
\end{scope}
\end{scope}
\begin{scope}
\draw[fill=white] (35.000000, 0.000000) circle(3.000000pt);
\clip (35.000000, 0.000000) circle(3.000000pt);
\draw (32.000000, 0.000000) -- (38.000000, 0.000000);
\draw (35.000000, -3.000000) -- (35.000000, 3.000000);
\end{scope}
\draw[color=black] (50.500000, 0.000000) node [fill=white] {$f$};
\draw (66.000000,39.000000) -- (66.000000,0.000000);
\begin{scope}[rounded corners=2pt]
\begin{scope}
\draw[fill=white] (66.000000, 26.000000) +(-45.000000:8.485281pt and 26.870058pt) -- +(45.000000:8.485281pt and 26.870058pt) -- +(135.000000:8.485281pt and 26.870058pt) -- +(225.000000:8.485281pt and 26.870058pt) -- cycle;
\clip (66.000000, 26.000000) +(-45.000000:8.485281pt and 26.870058pt) -- +(45.000000:8.485281pt and 26.870058pt) -- +(135.000000:8.485281pt and 26.870058pt) -- +(225.000000:8.485281pt and 26.870058pt) -- cycle;
\draw (66.000000, 26.000000) node {{$f_3$}};
\end{scope}
\end{scope}
\begin{scope}
\draw[fill=white] (66.000000, 0.000000) circle(3.000000pt);
\clip (66.000000, 0.000000) circle(3.000000pt);
\draw (63.000000, 0.000000) -- (69.000000, 0.000000);
\draw (66.000000, -3.000000) -- (66.000000, 3.000000);
\end{scope}
\draw (80.000000,39.000000) -- (80.000000,0.000000);
\begin{scope}[rounded corners=2pt]
\begin{scope}
\draw[fill=white] (80.000000, 26.000000) +(-45.000000:8.485281pt and 26.870058pt) -- +(45.000000:8.485281pt and 26.870058pt) -- +(135.000000:8.485281pt and 26.870058pt) -- +(225.000000:8.485281pt and 26.870058pt) -- cycle;
\clip (80.000000, 26.000000) +(-45.000000:8.485281pt and 26.870058pt) -- +(45.000000:8.485281pt and 26.870058pt) -- +(135.000000:8.485281pt and 26.870058pt) -- +(225.000000:8.485281pt and 26.870058pt) -- cycle;
\draw (80.000000, 26.000000) node {{$f_2$}};
\end{scope}
\end{scope}
\begin{scope}
\draw[fill=white] (80.000000, 0.000000) circle(3.000000pt);
\clip (80.000000, 0.000000) circle(3.000000pt);
\draw (77.000000, 0.000000) -- (83.000000, 0.000000);
\draw (80.000000, -3.000000) -- (80.000000, 3.000000);
\end{scope}
\draw (94.000000,39.000000) -- (94.000000,0.000000);
\begin{scope}[rounded corners=2pt]
\begin{scope}
\draw[fill=white] (94.000000, 26.000000) +(-45.000000:8.485281pt and 26.870058pt) -- +(45.000000:8.485281pt and 26.870058pt) -- +(135.000000:8.485281pt and 26.870058pt) -- +(225.000000:8.485281pt and 26.870058pt) -- cycle;
\clip (94.000000, 26.000000) +(-45.000000:8.485281pt and 26.870058pt) -- +(45.000000:8.485281pt and 26.870058pt) -- +(135.000000:8.485281pt and 26.870058pt) -- +(225.000000:8.485281pt and 26.870058pt) -- cycle;
\draw (94.000000, 26.000000) node {{$f_1$}};
\end{scope}
\end{scope}
\begin{scope}
\draw[fill=white] (94.000000, 0.000000) circle(3.000000pt);
\clip (94.000000, 0.000000) circle(3.000000pt);
\draw (91.000000, 0.000000) -- (97.000000, 0.000000);
\draw (94.000000, -3.000000) -- (94.000000, 3.000000);
\end{scope}
\draw[color=black] (101.000000,0.000000) node[right] {$0$};
\end{tikzpicture}}

  \caption{Mapping LUTs that represent the XOR function}
  \label{fig:xor-lut}
\end{figure}

We chose XAG graphs as logic representation, due to the inexpensive implementation of the XOR operation in fault tolerant quantum circuits. Following this idea we modify the mapping algorithm to identify and select cuts that performs the parity function. They can be synthesized as multiple-input XOR gates.  

After we perform cut enumeration we refine the cuts, looking for multi-input XOR blocks.
In fact, if all leaves of the cuts have a fan-out size of 1, i.e., they only fan-in into the XOR gate, we can apply an alternative mapping strategy that leads to reduction of qubits and gates.

Fig.~\ref{fig:xor-lut} illustrates the improved mapping strategy for
an XOR cut with three inputs.  In Fig.~\ref{fig:xor-lut}(a) this LUT
is drawn as an XOR symbol.  The conventional mapping,
shown in Fig.~\ref{fig:xor-lut}(b), maps each child in to a clean
ancilla, and then uses another clean ancilla to map the result of the
XOR cut.  The result of that cut, $f$, can then be used by its parents
in subsequent gates.  We illustrate this fact by simply annotating the
circuit line where it represents the value $f$.  However, since in
this case the child cuts $f_1$, $f_2$, and $f_3$ are composed via the
XOR operator, one can directly map them in to a single qubit without
the need of requiring an additional ancilla for each child LUT, see
Fig.~\ref{fig:xor-lut}(c).

Note that the size of the XOR gates does not need to be bounded by the
LUT size $k$. In order to build XOR blocks in XAGs, we
first detect 2-input XOR gates. Afterwards, sub-trees of XOR gates are grouped together.
Finally, we adjust cut enumeration such that XOR cuts are
assigned with cost 0, in order to force the LUT mapping to prefer XOR blocks.
\begin{table*}[t!]
\def\tabcolsep{4pt}
  \begin{tabularx}{\textwidth}{Xrrrrrrrrrr}
  \toprule
  &\multicolumn{2}{c}{M/B} & \multicolumn{2}{c}{S/B} & \multicolumn{2}{c}{S/P\_match\_q} & \multicolumn{2}{c}{S/P\_match\_g} & \multicolumn{2}{c}{M/P}\\
  \cmidrule(lr){2-3}
  \cmidrule(lr){4-5}
  \cmidrule(lr){6-7}
  \cmidrule(lr){8-9}
  \cmidrule(lr){10-11}
                 &  gates & qubits & gates  &qubits & gates & qubits & gates  & qubits  & gates & qubits\\
\midrule
addassoc4   &   1376  &    25   &    1029   &      34      &              1141          &            25          &         1371          &           22       &      1904      &         19 \\
addassoc5   &   2987  &    36   &    1586   &      49      &              1798          &            36          &         1804          &           31       &      5365      &         24 \\
addassoc6   &   2394  &    43   &    1445   &      58      &              1513          &            42          &         1729          &           35       &      8268      &         26 \\
addassoc7   &   3243  &    51   &    1941   &      70      &              2201          &            50          &         2361          &           44       &      4383      &         36 \\
addassoc8   &   3221  &    62   &    2018   &      79      &              2312          &            57          &         2430          &           49       &      4787      &         40 \\
addassoc9   &   3603  &    70   &    2385   &      89      &              2453          &            67          &         2773          &           56       &      5569      &         42 \\
addassoc10  &   4528  &    80   &    2835   &      97      &              3575          &            70          &         3549          &           58       &      6142      &         50 \\
multassoc4  &   6682  &    34   &    2751   &      60      &              3057          &            34          &         3193          &           33       &     10834      &         19 \\
multassoc5  &  10519  &    54   &    4811   &     104      &              5321          &            55          &           5321          &            55       &     16687      &         31 \\
multassoc6  &  17653  &    93   &    7395   &     172      &              8565          &            96          &            8565          &            96        &     22933      &         53 \\
multassoc7  &  25395  &   138   &   11099   &     240      &             15425          &           135          &        14607          &          128       &     37717      &         74 \\
multassoc8  &  32443  &   181   &   13781   &     323      &             20713          &           179          &        22997          &          166       &     51757      &         94 \\
multassoc9  &  37599  &   212   &   17881   &     394      &             34305          &           203          &        32489          &          200       &     66267      &        110 \\
multassoc10 &  47795  &   289   &   22843   &     525      &             41825          &           281          &        41081          &          262       &    101627      &        143 \\
multdistr4  &   4812  &    29   &    2368   &      54      &              3262          &            29          &         3694          &           25       &      5034      &         19 \\
multdistr5  &   9011  &    54   &    4569   &      94      &              5441          &            54          &         5441          &           54       &     22717      &         25 \\
multdistr6  &  13327  &    78   &    6092   &     143      &              7138          &            80          &           7138          &            80       &     15169      &         46 \\
multdistr7  &  18268  &   110   &    8771   &     200      &             13849          &           109          &        13849          &          109       &     21746      &         63 \\
multdistr8  &  26151  &   149   &   11888   &     276      &             17896          &           149          &        17520          &          143       &     39449      &         81 \\
multdistr9  &  30427  &   184   &   14477   &     332      &             22917          &           182          &        22445          &          181       &     43819      &         99 \\
multdistr10 &  37571  &   226   &   17808   &     414      &             29714          &           214          &        31570          &          219       &     55583      &        122 \\
\midrule
\multicolumn{5}{l}{S/P vs M/B average results} & -32.31\% & -1.86\%  & -29.77\% &  -8.38\% & \multicolumn{2}{r}{}\\
\bottomrule
  \end{tabularx}
  \caption{Comparison between \emph{ROS} and \emph{LHRS}}\label{table}
\end{table*}

\section{Applications and Results}\label{results}
We have implemented our algorithms into the hierarchical quantum synthesis framework caterpillar\footnote{https://github.com/gmeuli/caterpillar} in C++.
In this section we illustrate the efficiency of our proposed approach by synthesizing oracles, which can be used in algorithms such as Grover's search algorithm~\cite{Grover96}, which is capable of computing a satisfying assignment for a quantum oracle optimally with a quadratic speedup. 

For our benchmarks we suppose that we want to perform equivalence checking between two designs. 
Equivalence checking is a well-known problem in logic synthesis that has been addressed by many logic synthesis tools, as for example \textit{abc}~\cite{abc} or \textit{Formality}$\textregistered$.
We need to synthesize an oracle quantum circuit of the function $f$, where $f$ is satisfied when the two graphs performs a different operation. The algorithm would either prove that the two circuit are equivalent, or would provide the input set for which the two functions evaluate differently.

Our benchmark consists of XAG graphs. Each graph represents an equivalence checking miter of two circuits that perform the same function but using a different network structure. The miter of two networks is a network built by joining their input sets and by computing the 2-input XOR between their outputs. Further, one or more injected faults (a node performing a different computation) are injected in one of the two circuits.
We consider three type of benchmarks: \emph{addassoc}, where the algorithm should verify the validity of the associative property of addition; \emph{multassoc}, where the two designs should be equivalent thanks to the associativity of the multiplication, and \emph{multdistr}, to prove the distributivity of the multiplication.
Each benchmark is considered with bitwidths from $w = 4$ to $10$ bits. Consequently, each benchmark has $3w$ inputs and $1$ output.

Our experimental results are reported in Table~\ref{table}. The first two columns show the results of the state-of-the-art (M/B) synthesis flow, that uses a classic $k$-LUT mapper and the Bennett strategy to deal with garbage results (\emph{LHRS}). 

As expected, data shows that by only changing the $k$-LUT mapper (S/B) we always reduce the number of gates, paying in an increased number of qubits. On the other hand, by only applying the quantum garbage management technique (M/P), the number of qubits is always reduced, and the number of gates increased. 

In the S/P\_match\_q experiment we have used ROS, setting the number of qubits to match M/B. In most of the cases we obtain an improvement in both qubits and gates, with the exception of \textit{multassoc5}, \textit{multdistr6} and \textit{multassoc6}. For the latter cases, the SAT solver that is used in the quantum garbage management technique had reached our limit of 50000 conflicts. For this reason, we needed to slightly increase the number of qubits, still obtaining in all cases a reduction in gates with respect to M/B. 
\emph{ROS} in this setting reduces the number of gates of 32.31\% and the number of qubits of 1.86\% on average.

In the S/P\_match\_g experiment, we start from the results in S/P\_match\_q and try to beat them, by decrementing the number of qubits, as long as the number of gates does not exceed the one in M/B. Also here ROS manages to obtain better results than the state-of-the-art flow both in gates and qubits. Gates are reduced of 29.77\% while qubits are reduced of 8.38\% on average, with respect to M/B.

Most of the synthesis runs completed within a few seconds, none required more than one minute in the worst-case.

\section{Conclusion}
In this work we introduce ROS: a hierarchical quantum synthesis flow based on $k$-LUT networks. ROS exploits a $k$-LUT mapper that has been specifically designed for this application. This mapper is capable of generating LUTs that are easy to be synthesized by the Gray synthesis method, and leads to a quantum circuit with fewer gates if compared with existing mappers, i.e. \textit{mf} from \textit{abc}. In addition, ROS exploits a SAT-based quantum memory management technique to gain control over the number of qubits of the generated circuits. In our experiments we apply ROS for the synthesis of quantum oracles, which may be used in algorithms as the Grover's algorithm. Experimental results prove the ability of ROS to break the border of the pareto-point synthesis results, beating the existing framework in both qubits and number of gates.  

{\small

\subsubsection*{Acknowledgments}  This research was
supported by the Swiss National Science Foundation (200021-169084 MAJesty).
}%





\end{document}